\documentclass[a4paper,11pt]{article}
\usepackage{pos}
\usepackage{graphicx,psfrag,epsf,color}
\usepackage{amsmath,amsfonts}
\usepackage{physics}
\usepackage{siunitx}
\usepackage{subfig}

\title{CP Violation in $\boldmath{B \to K\ell^+\ell^-}$ Decays: New Opportunities in the High-Precision Era}

\author[a,b]{Robert Fleischer}
\author[a,c]{Eleftheria Malami}
\author*[a,b]{Anders Rehult}
\author[a,d]{K. Keri Vos}

\affiliation[a]{Nikhef, Science Park 105, NL-1098 XG Amsterdam,  Netherlands}

\affiliation[b]{Department of Physics and Astronomy, Vrije Universiteit Amsterdam, NL-1081 HV Amsterdam, Netherlands}

\affiliation[c]{Center for Particle Physics Siegen (CPPS), Theoretische Physik 1, Universität Siegen, D-57068 Siegen, Germany}

\affiliation[d]{Gravitational Waves and Fundamental Physics (GWFP), Maastricht University, Duboisdomein 30, NL-6229 GT Maastricht, the Netherlands}

\emailAdd{arehult@nikhef.nl}

\abstract{Experimental data on rare $B$-meson decays indicate deviations from Standard Model predictions. In studies of these decays, possible new sources of CP violation are often neglected. We discuss CP violation in the rare $B$-meson decays $B\to K\ell^+\ell^- (\ell = \mu,e)$ and point to two phenomena that arise when new sources of CP violation are included. First, the Wilson coefficients $C_{9\ell}$ and $C_{10\ell}$ become complex, and we show how we can extract their values from measurements of direct and mixing-induced CP asymmetries. Second, new sources of CP violation can generate nontrivial lepton flavour universality violation. Such a violation is usually measured through ratios like $R_K$ and $R_{K^*}$, but we show that measuring only these ratios leaves a large parameter space unexplored. These results bring exciting opportunities to reveal New Physics effects in the high-precision era.}

\FullConference{Corfu Summer Institute 2023 "School and Workshops on Elementary Particle Physics and Gravity" (CORFU2023)\\
 23 April - 6 May , and 27 August - 1 October, 2023\\
Corfu, Greece\\}

\begin{document}
\maketitle

\section{Introduction}
\noindent The decays $B \to K\ell^+\ell^- (\ell = \mu, e)$ belong to the class of rare $B$-meson decays, which in the Standard Model (SM) occur only at the loop level and are heavily suppressed. For this reason they are excellent probes of New Physics (NP). Rare $B$-meson decays have recently received a lot of attention in the literature (see e.g. \cite{Bobeth:2017vxj,Alok:2017sui, Datta:2019zca,Altmannshofer:2021qrr,Alguero:2022wkd,Gubernari:2022hxn,Geng:2021nhg,Carvunis:2021jga,Mahmoudi:2022hzx,SinghChundawat:2022zdf}), much of it owing to LHCb’s observation that the branching ratio of $B \to K\mu^+\mu^-$ is lower than the SM prediction, with a statistical significance of $ \sim 4 \sigma$ \cite{LHCb:2014cxe,Fleischer:2022klb}.

If this deviation is confirmed to be caused by NP, the next step is to identify the nature of that NP. In the language of effective field theory, NP is encoded in the Wilson coefficients that characterize the given decay. These coefficients provide a model-independent way to describe rare $B$-meson decays, and they can be calculated in different high-energy theories. If we can extract their values from data, we can see which high-energy theories can explain the observed deviation. Wilson coefficients are often assumed to be real numbers, but new sources of CP violation beyond the SM would make them complex. Following \cite{Fleischer:2022klb}, we will show how measurements of CP asymmetries in $B^\pm \to K^\pm \mu^+ \mu^-$ and $B^0_d \to K_S \mu^+ \mu^-$ decays can be used to extract the complex values of the relevant Wilson coefficients.

A particularly interesting question is whether lepton flavour universality is violated in rare $B$-meson decays. This question has been explored through measurements of the ratio \cite{Hiller:2003js}
\begin{equation}
R_K  = \frac{ \Gamma(B^-\to K^-\mu^+\mu^-) +\Gamma(B^+\to K^+\mu^+\mu^-) }{\Gamma(B^-\to K^-e^+e^-)  + \Gamma(B^+\to K^+e^+e^-) } ,
\label{eq:RK}
\end{equation}
which probes universality between muons and electrons. For several years, measurements of this ratio deviated from the SM value of unity \cite{LHCb:2014vgu,LHCb:2019hip,LHCb:2021trn,Bordone:2016gaq}. However, as of December 2022, the value is in agreement with the SM prediction within the current uncertainties \cite{LHCb:2022qnv,LHCb:2022zom, Isidori:2022bzw}. Does this mean that electron--muon universality is now tightly constrained in $B \to K\ell^+\ell^-$? As we will show, following \cite{Fleischer:2023zeo}, that is not the case: With new CP-violating phases entering the decays, we can still have electron--muon non-universality even with $R_K \sim 1$. We will discuss CP-violating observables that offer exciting new perspectives at the high-precision frontier.

\section{Theoretical framework}
\label{ch:B_to_Kellell_theoretical_framework}
\subsection{Effective Hamiltonian}
\noindent For $b \to s \ell^+\ell^-$ transitions, the relevant low-energy effective Hamiltonian is given as follows \cite{Descotes-Genon:2020tnz, Altmannshofer:2008dz, Gratrex:2015hna,Buchalla:1995vs}:
\begin{equation}\label{eq:ham}
    \mathcal{H}_{\rm eff} = - \frac{4 G_F}{\sqrt{2}} \left[\lambda_u \Big\{C_1 (\mathcal{O}_1^c - \mathcal{O}_1^u) + C_2 (\mathcal{O}_2^c - \mathcal{O}_2^u)\Big\} + \lambda_t \sum\limits_{i \in I} C_i \mathcal{O}_i \right] \ ,
\end{equation}
where the label $I = \{1c, 2c, 3, 4, 5, 6, 8, 7^{(\prime)}, 9^{(\prime)}\ell, 10^{(\prime)}\ell, S^{(\prime)}\ell, P^{(\prime)}\ell, T^{(\prime)}\ell\}$ distinguishes different operators and $\lambda_q = V_{qb} V_{qs}^*$ are CKM factors. We neglect the doubly Cabibbo-suppressed terms proportional to $\lambda_u$, which contribute at the $\mathcal{O}(\lambda^2) \sim 5\%$ level, and focus on the following operators:
\begin{equation}
    \mathcal{O}_{9^{(\prime)}\ell} = \frac{e^2}{(4\pi)^2} [\bar s \gamma^\mu P_{L(R)} b] (\bar \ell \gamma_\mu \ell), \quad \quad 
    \mathcal{O}_{10^{(\prime)}\ell} = \frac{e^2}{(4\pi)^2} [\bar s \gamma^\mu P_{L(R)} b] (\bar \ell \gamma_\mu \gamma_5 \ell) \ ,
\end{equation}
with $P_{R(L)} = \frac{1}{2} (1 \pm \gamma_5)$. The index $\ell = \mu, e$ indicates lepton flavour, and we will suppress that label when it is clear from context that a specific flavour is considered. Furthermore, all Wilson coefficients $C_i$ should in the rest of this write-up be understood as a shorthand notation for $C_i + C_i^\prime$.

\subsection{Observables}
\noindent We will work with three observables: Branching ratios, direct CP asymmetries, and mixing-induced CP asymmetries. For detailed expressions, see \cite{Fleischer:2022klb}. We define the $q^2$-integrated direct CP asymmetry of $B^\pm \to K^\pm \mu^+\mu^-$ as 
\begin{equation}
    \mathcal{A}_{\rm CP}^{\rm dir}[q^2_{\rm min}, q^2_{\rm max}] = \frac{\bar \Gamma[q^2_{\rm min}, q^2_{\rm max}] - \Gamma[q^2_{\rm min}, q^2_{\rm max}]}{\bar \Gamma[q^2_{\rm min}, q^2_{\rm max}] + \Gamma[q^2_{\rm min}, q^2_{\rm max}]} \ ,
    \label{eq:q2_binned_CP_asymm}
\end{equation}
where $\bar \Gamma \equiv \Gamma(B^- \to K^- \ell^+\ell^-)$ and $\Gamma \equiv \Gamma(B^+ \to K^+ \ell^+\ell^-)$ with
\begin{equation}\label{eq:brdef}
    \Gamma[q^2_{\rm min}, q^2_{\rm max}] = \int_{q^2_{\rm min}}^{q^2_{\rm max}} \frac{d\Gamma}{dq^2} dq^2 \ .
\end{equation}

In contrast to charged $B$-meson decays, which have only direct CP violation, neutral $B$-meson decays can also show mixing-induced CP violation. This phenomenon arises through $B^0_q$--$\bar B^0_q$ oscillations ($q \in \{d,s\}$) if the $B^0_q$ and $\bar B^0_q$ mesons can both decay into the same final state. Here, we consider the decay $B^0_d \to K_S\ell^+\ell^-$. We define its mixing-induced CP asymmetry through the following time-dependent decay rate \cite{Fleischer:2017yox}:
\begin{equation}
\begin{aligned}
    \frac{\Gamma(B^0_d(t) \to K_S \ell^+\ell^-) 
    - \Gamma(\bar B^0_d(t) \to K_S \ell^+\ell^-)}{\Gamma(B^0_d(t) \to K_S \ell^+\ell^-)
    + \Gamma(\bar B^0_d(t) \to K_S \ell^+\ell^-)} 
    &=\frac{- \mathcal{A}_{\rm CP}^{\rm dir}\cos(\Delta M_d t)  -\mathcal{A}_{\rm CP}^{\rm mix} \sin (\Delta M_d t) }{\cosh (\frac{\Delta \Gamma_d}{2} t) + \mathcal{A}_{\Delta \Gamma} \sinh (\frac{\Delta \Gamma_d}{2} t)} \ ,
\end{aligned}
\label{eq:time_dependent_CP_asymmetry}
\end{equation}
where $\mathcal{A}_{\rm CP}^{\rm mix}$ is the mixing-induced CP asymmetry and $\mathcal{A}_{\rm CP}^{\rm dir}$ is the direct CP asymmetry of~\eqref{eq:q2_binned_CP_asymm}. Here, $\Delta \Gamma_{q} = \Gamma_H^d - \Gamma_L^d$ is decay width difference between the heavy and light $B_d$ mass eigenstates, and $\Delta M_d = M_H^d - M_L^d$ is their mass difference.

The direct and mixing-induced CP asymmetries can be affected by complex phases in $C_{9\ell}$ and/or $C_{10\ell}$. The two asymmetries have complementary sensitivities to these coefficients: The direct asymmetry depends only $C_{9\ell}$, while the mixing-induced one depends on both $C_{9\ell}$ and $C_{10\ell}$. We will now use this fact to extract the CP-violating phases of the two coefficients.

\section{\boldmath Extracting complex Wilson coefficients from CP asymmetries}
\label{ch:fingerprinting}
\noindent In this section, based on \cite{Fleischer:2022klb}, we show how CP asymmetries in $B \to K\mu^+\mu^-$ decays allow us to extract the complex coefficients $C_{9\mu}$ and $C_{10\mu}$. We first discuss the relevant experimental bounds, then demonstrate how the CP-violating observables give us complementary information, and finally extract the values of $C_{9\mu}$ and $C_{10\mu}$ in a fit to hypothetical data from a future benchmark scenario.

\subsection{Experimental bounds}
\label{ch:exp_bounds}
\begin{figure}
    \centering
    \includegraphics[width=0.5\textwidth]{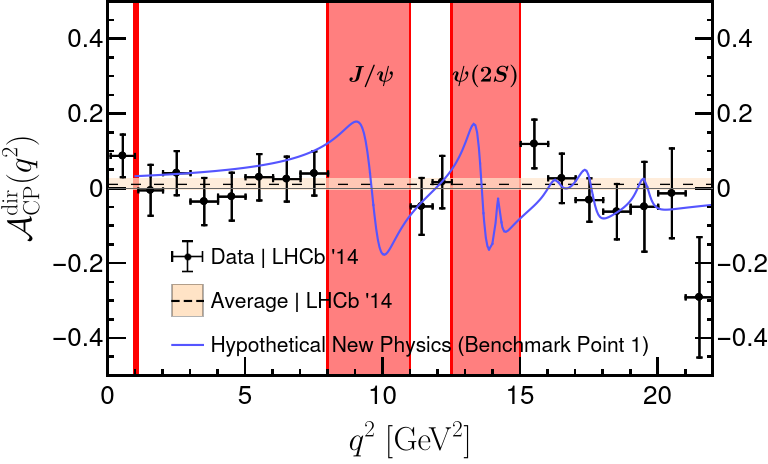}
    \caption{Experimental data on the direct CP asymmetry of $B^- \to K^- \mu^+ \mu^-$ from the LHCb collaboration \cite{LHCb:2014mit}. The blue line indicates a theory prediction following Benchmark Point 1 of~\eqref{eq:staranddiamond}. (From \cite{Fleischer:2022klb}. Similar figures can be found in \cite{Becirevic:2020ssj,Smolkovic:2022axy}.)}
    \label{fig:ACP_bounds}
\end{figure}

\noindent The LHCb collaboration has measured the branching ratio and direct CP asymmetry of the decay $B^\pm \to K^\pm \mu^+\mu^-$. We will begin by focusing on the $q^2$ bin of $[7,8] \; \si{\giga eV^2}$, where LHCb finds \cite{LHCb:2014cxe}
\begin{equation}
    \mathcal{B}(B^+ \to K^+ \mu^+\mu^-)[7,8] = (23.1 \pm 1.8) \times 10^{-9}
\label{eq:expBR_7to8}
\end{equation}
and \cite{LHCb:2014mit}
\begin{equation}\label{eq:adirmeas}
        \mathcal{A}_{\rm CP}^{\rm dir}[7,8] = 0.041 \pm 0.059 \ .
\end{equation}
In Fig.~\ref{fig:ACP_bounds}, we show data on the direct CP asymmetry in different $q^2$-bins. We focus on the small bin of $[7,8] \; \si{\giga \eV^2}$ because it is near the $J/\psi$ resonance at $m_{J/\psi}^2 = 9.6 \; \si{\giga\eV^2}$, where the direct CP asymmetry could be enhanced \cite{Becirevic:2020ssj}. Concerning the mixing-induced asymmetry, there are not yet any data available. In \cite{Fleischer:2022klb}, we provide SM predictions of all these observables.

For illustration, we consider three NP scenarios, which all fit the current rare $B$-meson decay data better than the SM \cite{Altmannshofer:2021qrr,Geng:2021nhg,Carvunis:2021jga,Mahmoudi:2022hzx,SinghChundawat:2022zdf}:
\begin{equation}
\begin{aligned}
    &\text{Scenario 1:} \quad C_9^{\rm NP} \neq 0 \ ,\\
    &\text{Scenario 2:} \quad C_9^{\rm NP} = -C_{10}^{\rm NP} \neq 0 \ ,\\
    &\text{Scenario 3:} \quad C_{10}^{\rm NP} \neq 0 \ .
\end{aligned}
\label{eq:NP_ranges}
\end{equation}

\begin{figure}
    \centering
    \subfloat[$C_9^{\rm NP}$ only]{\includegraphics[width=0.3\textwidth]{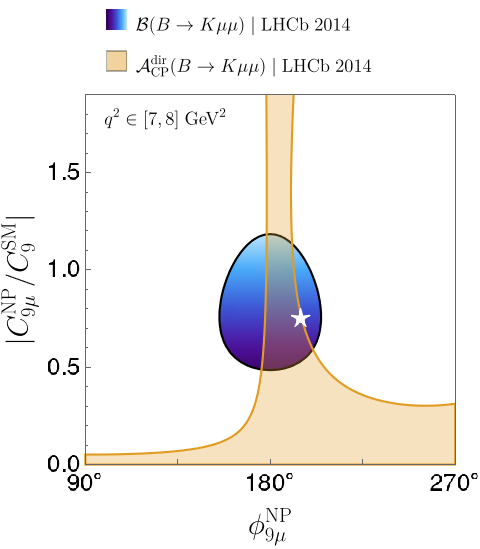}}
    \hfill
    \subfloat[$C_9^{\rm NP} = -C_{10}^{\rm NP}$]{\includegraphics[width=0.3\textwidth]{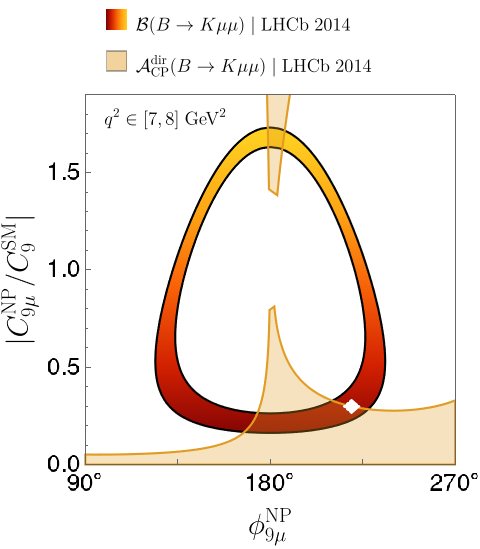}}
    \hfill
   \subfloat[$C_{10}^{\rm NP}$ only]{\includegraphics[width=0.3\textwidth]{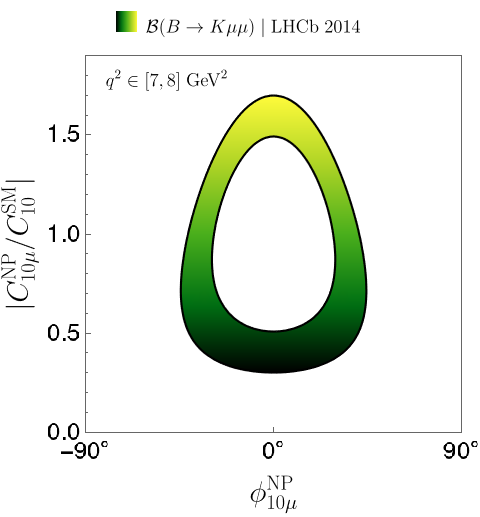}}
    \caption{Experimental $1\sigma$ bounds on the three NP scenarios of~\eqref{eq:NP_ranges}. The star and diamond indicate Benchmark Points 1 and 2 of~\eqref{eq:staranddiamond}, respectively. The colour gradients indicate the magnitudes of the Wilson coefficients. (From \cite{Fleischer:2022klb}.)
    }
    \label{fig:C9mu_bounds_scenario_1}
\end{figure}

Fig.~\ref{fig:C9mu_bounds_scenario_1} shows the $1 \sigma$ experimental bounds on the complex Wilson coefficients of these three scenarios. The oval regions are bounds from the branching ratio, while the light orange ones come from the direct CP asymmetry. The colour gradients indicate the magnitudes of the NP contributions. For these plots, we have used a CKM input following \cite{DeBruyn:2022zhw} that combines a value of $|V_{ub}|$ from exclusive semileptonic $B$ decays with a value of $|V_{cb}|$ from inclusive ones. We have also assumed a model of hadronic long-distance effects from \cite{LHCb:2016due} (see \cite{Fleischer:2022klb} for a detailed discussion). We have indicated by a star and diamond two benchmark points which are allowed by the data but still have large CP-violating phases: 
\begin{equation}
\begin{aligned}
    \text{Benchmark Point 1:}& \quad \abs{C_9^{\rm NP}}/\abs{C_9^{\rm SM}} = 0.75 \ , \quad\quad  \phi_9^{\rm NP} = 195^\circ \ ,\\
    \text{Benchmark Point 2:}& \quad \abs{C_9^{\rm NP}}/\abs{C_9^{\rm SM}} = \abs{C_{10}^{\rm NP}}/\abs{C_9^{\rm SM}}= 0.30\quad \phi_9^{\rm NP} = \phi_{10}^{\rm NP} - \pi = 220^\circ \ .
\end{aligned}
\label{eq:staranddiamond}
\end{equation}
The blue line in Fig.~\ref{fig:ACP_bounds} shows how Benchmark Point 1 leads to an enhanced CP asymmetry near $c \bar c$ resonances, following \cite{Becirevic:2020ssj}. This enhancement also occurs for Benchmark Point 2, but to a smaller extent. We will use these points in Section~\ref{ch:extracting_WCs}.

\subsection{Correlations between CP-violating observables}
\label{ch:correlations_one_branch}
\begin{figure}
    \centering
    \includegraphics[width=0.6\textwidth]{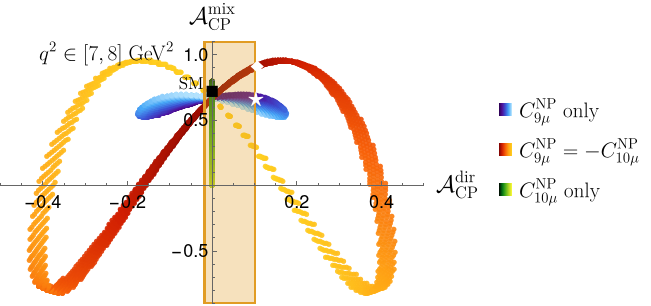}
    \caption{Correlations between direct and mixing-induced CP asymmetries in $B_d \to K_S\mu^+\mu^-$ in the three NP scenarios of~\eqref{eq:NP_ranges}. The orange vertical band marks the current experimental bound on the direct CP asymmetry. The colours for each scenario correspond to Fig. \ref{fig:C9mu_bounds_scenario_1} and indicate magnitudes of Wilson coefficients: Lighter colours indicate larger magnitudes. (From \cite{Fleischer:2022klb}.)}
\label{fig:Adir_Amix_correlation_toy_scenario}
\end{figure}

\noindent Through correlations between the direct and mixing-induced CP asymmetries of $B_d \to K_S \mu^+\mu^-$, we can distinguish between the three NP scenarios of~\eqref{eq:NP_ranges}. Mapping the experimentally allowed regions of Fig.~\ref{fig:C9mu_bounds_scenario_1} onto the $(\mathcal{A}_{\rm CP}^{\rm dir},\mathcal{A}_{\rm CP}^{\rm mix})$ plane, we obtain the correlations of Fig.~\ref{fig:Adir_Amix_correlation_toy_scenario}. In this plane, each NP scenario leaves a distinct ``fingerprint'', with different allowed values. In Scenario 1, the direct CP asymmetry can take any value within $[-0.2, 0.2]$ while the mixing-induced CP asymmetry stays close to the SM prediction. In Scenario 2, we get much larger asymmetries, with direct CP asymmetries as large as $\pm 0.4$ and mixing-induced ones in $[-0.8,0.9]$. In Scenario 3, the direct asymmetry is always zero, while the mixing-induced asymmetry varies within $[0, 0.8]$. The strikingly different behavior of the CP asymmetries between the NP scenarios demonstrates that these observables can be used to distinguish between them.

\subsection{Extracting Wilson coefficients from the CP asymmetries}
\label{ch:extracting_WCs}
\noindent We now go one step further and show how CP-violating observables in $B_d \to K_S \mu^+\mu^-$ allow us to extract the complex values of the Wilson coefficients $C_9$ and $C_{10}$. As the coefficients are complex numbers, they have two degrees of freedom each, and hence we need at least four observables to extract their values. For this analysis, we consider a minimal scenario with four observables, but we stress that additional observables would help overconstrain the system and produce a better fit. We consider:
\begin{itemize}
\item the CP-averaged branching ratio of $B^\pm \to K^\pm \mu^+\mu^-$ in the two $q^2$ bins of $q^2 \in [1.1,6.0]$ and $q^2 \in [15,22]$ \ ,
\item the direct CP asymmetry of $B^\pm \to K^\pm \mu^+\mu^-$ in the $q^2$ bin of $[8,9]$ \ ,
\item the mixing-induced CP asymmetry of $B^0_d \to K_S \mu^+\mu^-$ in the $q^2$ bin of $[1.1,6.0]$ \ .
\end{itemize}
These choices of bins play to the strengths of each observable. The direct CP asymmetry is chosen close to a $c \bar c$ resonance, where it can be maximally enhanced, while the branching ratios and mixing-induced CP asymmetry are chosen outside the resonance region, where they are robust with respect to uncertainties from hadronic long-distance effects.

\begin{figure}
    \centering
    \includegraphics[width=0.37\textwidth]{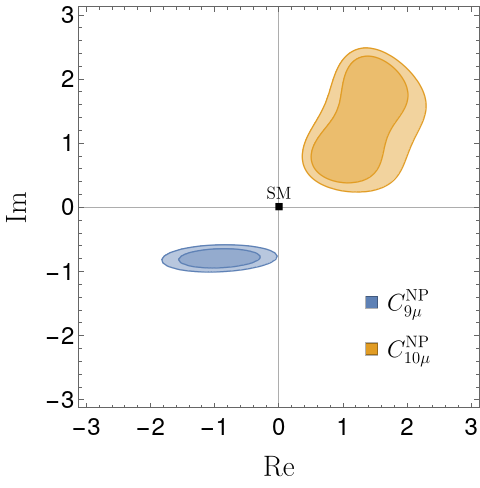}
    \caption{Extracted values of $C_9$ and $C_{10}$ in the complex plane for a future scenario discussed in the text. The lines indicate the $68\%$ and $90\%$ confidence limits. (From \cite{Fleischer:2022klb}.)}
    \label{fig:fitresults}
\end{figure}

For illustration, we consider Benchmark Point 2 of~\eqref{eq:staranddiamond}. Using this point to compute the values of our observables, we get 
\begin{align}\label{eq:input}
\mathcal{A}_{\rm CP}^{\rm dir}[8,9] &= 0.16 \pm 0.02\ , & \mathcal{A}_{\rm CP}^{\rm mix}[1.1,6] &= 0.94 \pm 0.04 \ ,  \\
\mathcal{B}[1.1,6.0] &= (1.15\pm 0.02)\times 10^{-7} \ , &  \mathcal{B}[15,22] &= (0.908\pm 0.018)\times 10^{-7} \nonumber \ ,
\end{align}
where the uncertainties indicate a hypothetical future experimental scenario. We have here chosen the uncertainty on $\mathcal{A}_{\rm CP}^{\rm dir}$ to be about one third of the current experimental uncertainty in the nearest available $q^2$ bins \cite{Aaij_2014_direct_CPV}. For $\mathcal{A}_{\rm CP}^{\rm mix}$, we have chosen an uncertainy twice the size of that on the direct CP asymmetry. Finally, for the uncertainties on the branching ratios, we have assumed $2 \%$, again about one third of current experimental uncertainties \cite{LHCb:2014cxe}. Working with these uncertainties lets us explore how a given precision on observables translates into precision on the extracted Wilson coefficients. We do this by performing a chi-squared fit while setting theory uncertainties to zero. Fig.~\ref{fig:fitresults} shows the resulting $68
\%$ and $90\%$ confidence level regions for the Wilson coefficients in the complex plane. We find that the corresponding observables allow us to determine the imaginary part of $C_9$ with high precision, but we get a less precise determination of $C_{10}$. The precision can be increased by over-constraining the system, e.g. by including additional $q^2$ bins, by considering new observables, or by assuming relations between the Wilson coefficients.
\begin{figure}
    \centering
    {\includegraphics[width=0.48\textwidth]{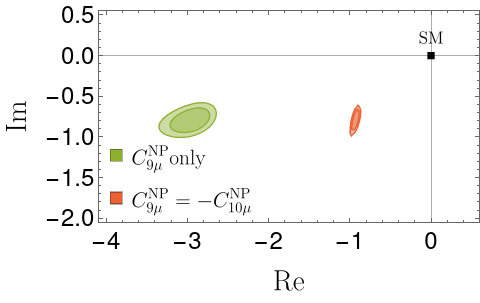}}
    \caption{Result from an overconstrained fit. The green region comes from assuming NP in $C_9$ only, while the red region comes from assuming $C_9 = -C_{10}$.}
    \label{fig:fitresults_b}
\end{figure}

To show the precision gained from an over-constrained fit, we consider two scenarios with NP either in $C_9$ only or in $C_9=-C_{10}$. For the $C_9$-only scenario, we use Benchmark Point 1 from~\eqref{eq:staranddiamond} and obtain
\begin{align}\label{eq:input_scenario1}
\mathcal{A}_{\rm CP}^{\rm dir}[8,9] &= 0.15 \pm 0.02\ , & \mathcal{A}_{\rm CP}^{\rm mix}[1.1,6] &= 0.66 \pm0.04 \ ,  \\
\mathcal{B}[1.1,6.0] &= (1.16\pm 0.02)\times 10^{-7} \ , &  \mathcal{B}[15,22] &= (0.806\pm 0.016)\times 10^{-7} \nonumber \ ,
\end{align}
where the uncertainties again correspond to a hypothetical future scenario. With these inputs for $C_9$ only and those of~\eqref{eq:input} for $C_9=-C_{10}$, we find the more precise confidence limits of Fig.~\ref{fig:fitresults_b}.

\section{Testing electron--muon universality with CP violation}
\label{sec:tests}
\noindent With current experimental data on $R_K$ in \eqref{eq:RK} being in agreement with the SM prediction, is there still space left for electron--muon universality violation? If the Wilson coefficients in $B \to K\mu^+\mu^-$ and $B \to Ke^+e^-$ are the same, then we have electron--muon universality. However, as we will show in this section following \cite{Fleischer:2023zeo}, when we allow the coefficients to be complex, they can take different values while still respecting $R_K \sim 1$. We will first consider real coefficients and then complex ones, allowing for new sources of CP violation.

\subsection{Real Wilson coefficients}
\begin{figure}
    \centering
    \includegraphics[width=0.28\textwidth]{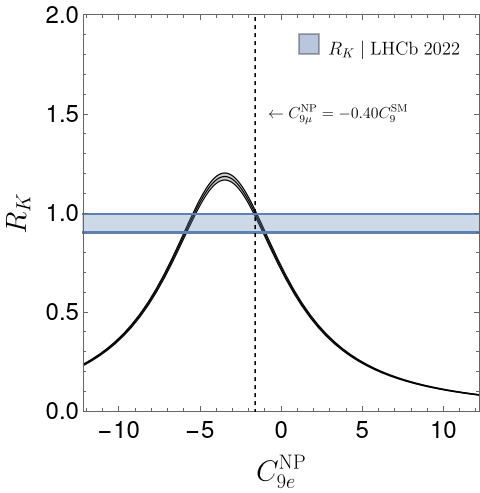}
    \caption{$R_K$ as a function of a real $C_{9e}^{\rm NP}$, corresponding to no new sources of CP violation. (From \cite{Fleischer:2023zeo}.)}
    \label{fig:fig1}
\end{figure}

\noindent For illustration, we consider a scenario with NP only in $C_{9\mu}$ and $C_{9e}$. To accommodate data on the branching ratio $\mathcal{B}(B^+ \to K^+\mu^+\mu^-)$ \cite{LHCb:2014cxe}, a real $C_{9\mu}^{\rm NP}$ has to take a value within 
\begin{equation}
\begin{aligned}\label{eq:range}
\frac{C_{9\mu}^{\rm NP}}{C_9^{\rm SM}} = [-1.32, -0.40] \ .
\end{aligned}
\end{equation}
We fix $C_{9\mu}^{\rm NP}$ to a value within this range and then use the recent $R_K$ measurement to compute what values are allowed for $C_{9e}^{\rm NP}$. Fig.~\ref{fig:fig1} shows the resulting plot, where the dashed vertical line is the value that $C_{9\mu}^{\rm NP}$ is fixed to, the curve is $R_K$ as a function of $C_{9e}^{\rm NP}$, and the horizontal band is the experimental $1 \sigma$ band for $R_K$. We observe that $C_{9e}$ is forced to take one of two discrete values, one conserving and one violating electron--muon universality. As a result, if we assume real Wilson coefficients, the recent $R_K$ data impose electron--muon universality up to a twofold ambiguity.

\subsection{Complex Wilson coefficients}
\begin{figure}
    \centering
    \includegraphics[width=0.7\textwidth]{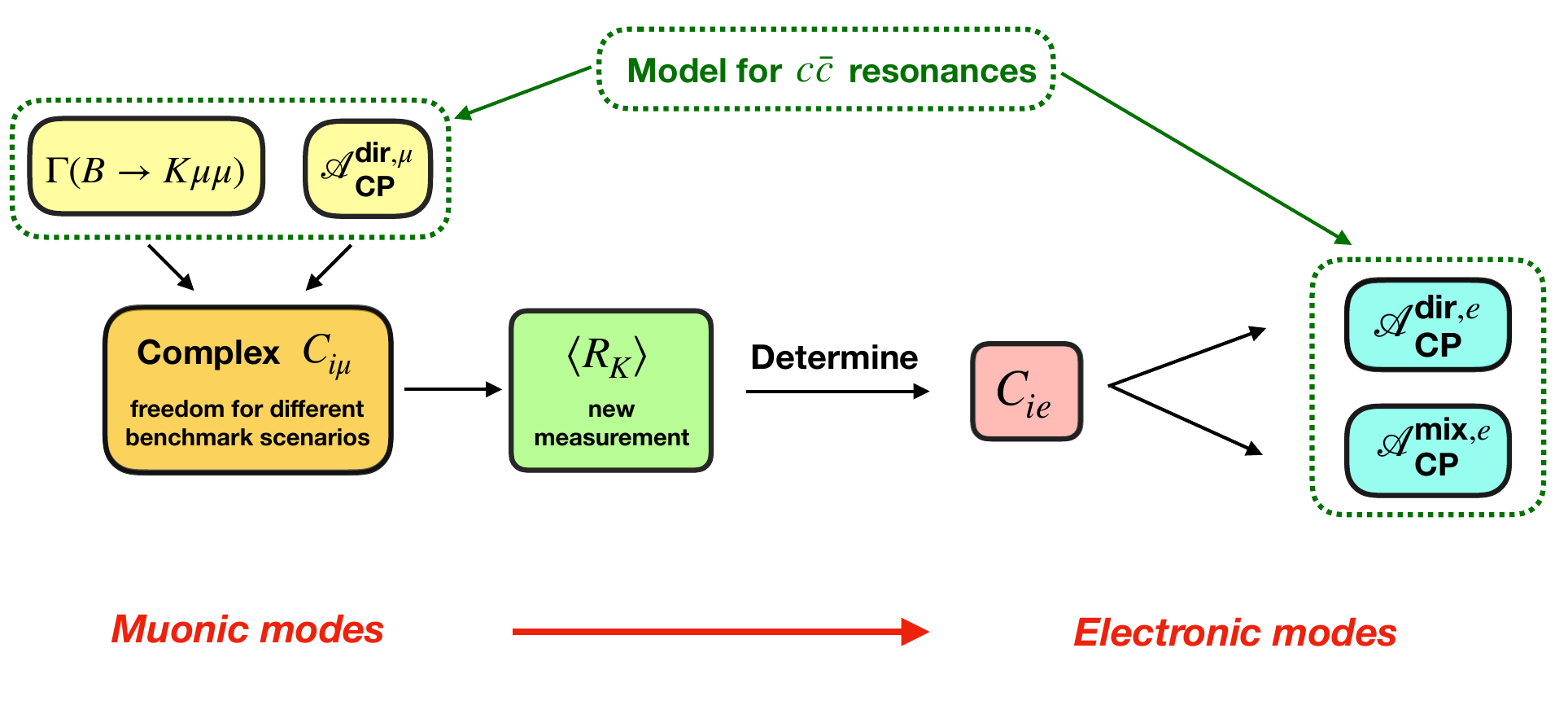}
    \caption{Illustration of our procedure to explore how much space is left for electron--muon non-universality in $B \to K\ell^+\ell^-$ when we allow for new sources of CP violation. (From \cite{Fleischer:2023zeo}.)}
    \label{fig:flowchart}
\end{figure}
\begin{figure}
    \centering
    \subfloat[]{\includegraphics[width=0.27\textwidth]{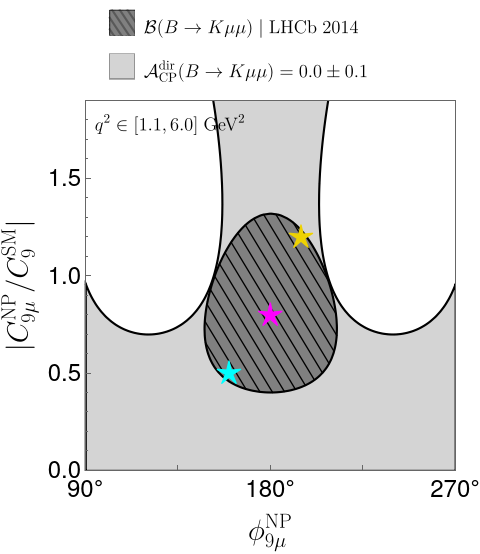} \label{fig:fig2a}}
    \subfloat[]{\includegraphics[width=0.27\textwidth]{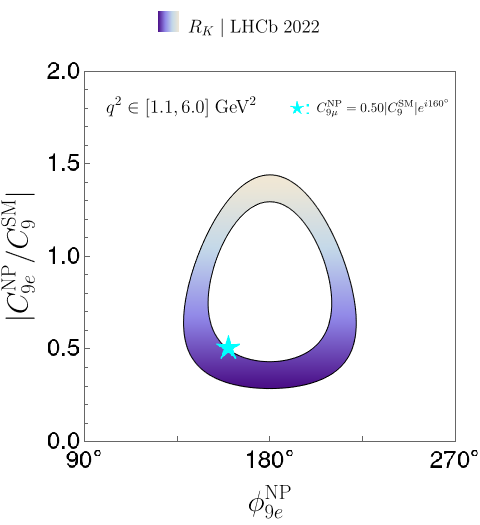} \label{fig:fig2b}}
    \subfloat[]{\includegraphics[width=0.27\textwidth]{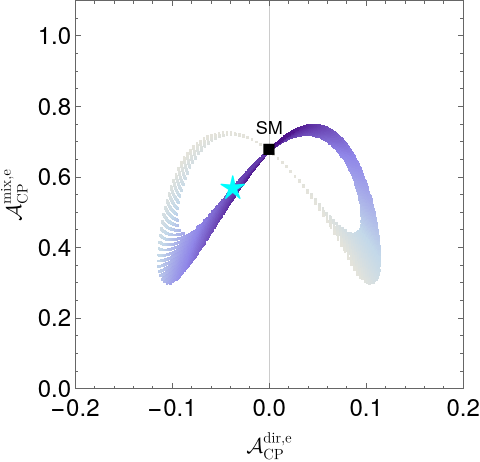} \label{fig:fig2c}}
    \caption{Constraints on the complex coefficients $C_{9\mu}^{\rm NP}$ (left), $C_{9e}^{\rm NP}$ (middle), and CP asymmetries in $B_d \to K_S e^+e^-$ (right). (From \cite{Fleischer:2023zeo}.)}
\label{fig:fig2}
\end{figure}
\begin{figure}
    \centering
    \subfloat[]{\includegraphics[width=0.27\textwidth]{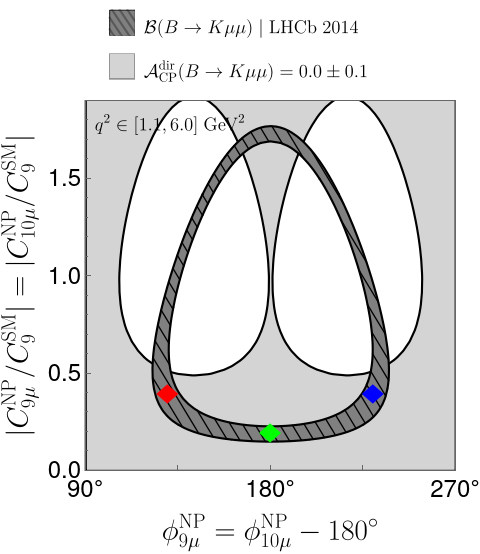}}
    \subfloat[]{\includegraphics[width=0.27\textwidth]{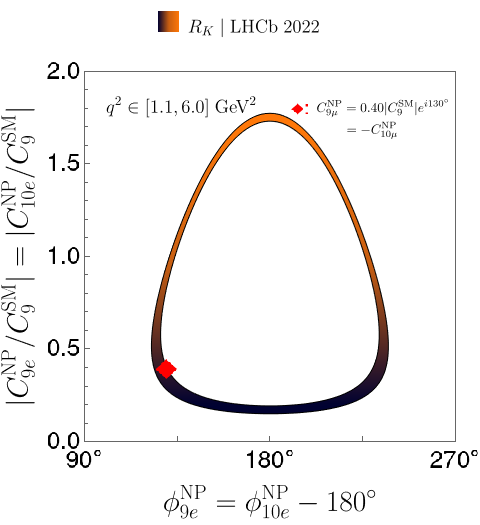}}
    \subfloat[]{\includegraphics[width=0.27\textwidth]{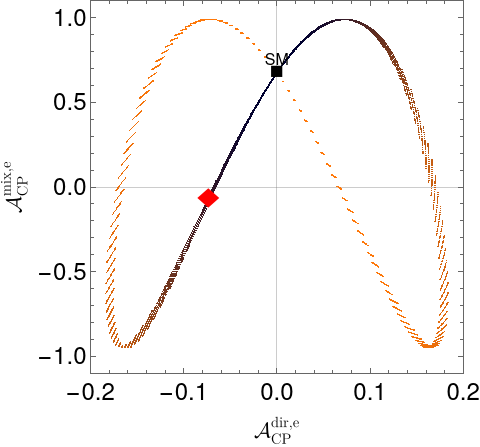}}
    \caption{Same as Fig.~\ref{fig:fig2} but for the NP scenario $C_{9\ell}^{\rm NP} = -C_{10\ell}^{\rm NP}$. (From \cite{Fleischer:2023zeo}.)}
\label{fig:fig2_scenario2}
\end{figure}
\noindent We now consider $C_{9\mu}^{\rm NP}$ and $C_{9e}^{\rm NP}$ to be complex parameters, allowing for new sources of CP violation. Fig.~\ref{fig:flowchart} schematically illustrates our procedure: we first constrain $C_{9\mu}^{\rm NP}$ using the branching ratio and direct CP asymmetry of $B^\pm \to K^\pm \mu^+\mu^-$, then use the current $R_K$ data to determine the allowed complex values of $C_{9e}^{\rm NP}$, and finally show how these values translate into CP-violating observables. Fig.~\ref{fig:fig2a} illustrates the experimental bounds on $C_{9\mu}^{\rm NP}$: The dark striped egg shows the bound from $\mathcal{B}(B^\pm \to K^\pm \mu^+\mu^-)$, and the light gray region shows the bound from $\mathcal{A}_{\rm CP}^{\rm dir}(B^\pm \to K^\pm \mu^+\mu^-)$. The direct CP asymmetry has been constrained by the LHCb collaboration in different $q^2$ bins \cite{LHCb:2014mit}. Averaging over the bins yields the rather strict bound of $\mathcal{A}_{\rm CP}^{\rm dir}(B\to K\mu^+\mu^-) = 0.012 \pm 0.017$. However, by taking this average, we would implicitly assume that the CP asymmetry is constant over the $q^2$ spectrum. As shown in \cite{Becirevic:2020ssj}, this is not necessarily the case: Any existing asymmetry will be enhanced by $c \bar c$ resonances and change across the spectrum. Because of this, we instead use the more conservative range of
\begin{equation}
    \mathcal{A}_{\rm CP}^{\rm dir}(B^-\to K^-\mu^+\mu^-) = 0.0 \pm 0.1 \ ,
\end{equation}
where we take the uncertainty to cover all the individual bins within $[1.1, 6.0] \; \si{\giga\eV^2}$ in \cite{LHCb:2014mit}.

For illustration, we fix $C_{9\mu}$ to the blue star, given by
\begin{equation}
    C_{9\mu}^{\rm NP} = 0.50 \abs{C_9^{\rm SM}} e^{i 160^\circ} \ ,
\end{equation}
and vary the electronic coefficient $C_{9e}^{\rm NP}$ to explore which values are consistent with $R_K$. Fig.~\ref{fig:fig2b} shows the result. We have coloured each point in the plot according to the absolute value of $C_{9e}^{\rm NP}$. If electron--muon universality were to hold in $B \to K\ell^+\ell^-$, then $C_{9e}^{\rm NP}$ should take the same value as $C_{9\mu}^{\rm NP}$, i.e. the blue star. But that is not what the figure shows. Instead, the electronic coefficient can take any value within the blue, egg-shaped region. Even with a value of $R_K$ close to one, there is still a lot of space left for electron--muon non-universality.

To probe this new parameter space, we need measurements of direct and mixing-induced CP asymmetries in $B_d \to K_S \mu^+\mu^-$ and $B_d \to K_S e^+e^-$. Fig.~\ref{fig:fig2c} shows how the allowed space in the $C_{9e}^{\rm NP}$ complex plane translates into the electronic CP asymmetry plane. The colour coding matches that of Fig.~\ref{fig:fig2b} and allows for easy comparison between the two planes. With data on either asymmetry, we can draw a band in the CP asymmetry plane, and with data on both we can draw two intersecting bands. If that intersection were to exclude a known value of $C_{9\mu}^{\rm NP}$ (in this case the blue star), it would constitute a clear signal of electron--muon universality violation.

The only data available on CP violation in $B \to Ke^+e^-$ come from the Belle Collaboration, which measured the direct CP asymmetry of $B^\pm \to K^\pm e^+e^-$ and found \cite{Belle:2009zue}:
\begin{equation}\label{eq:Adir_e}
    \mathcal{A}_{\rm CP}^{\rm dir, e} = 0.14 \pm 0.14 \ , 
\end{equation}
which is an average over different $q^2$ bins. Once measurements become available of the corresponding mixing-induced asymmetry of $B_d \to K_S e^+e^-$, we can use the method presented in \cite{Fleischer:2022klb} to determine the complex value of $C_{9e}^{\rm NP}$. Then, we would know whether electron--muon universality holds in these decays. Considering the significant room left for CP-violating couplings that violate electron--muon universality, we encourage the experimental community to perform detailed feasibility studies of the corresponding measurements.

We stress that these results are robust with respect to the specific benchmark value chosen for $C_{9\mu}^{\rm NP}$ and the particular NP scenario considered. To demonstrate this, we show in Fig.~\ref{fig:fig2_scenario2} the same plots as in Fig.~\ref{fig:fig2} but for the NP scenario of $C_{9\ell}^{\rm NP} = -C_{10\ell}^{\rm NP}$. Here, the red diamond benchmark point is given by
\begin{equation}
    C_{9\mu}^{\rm NP} = -C_{10\mu}^{\rm NP}= 0.40 \abs{C_9^{\rm SM}} e^{i 130^\circ} \ .
\end{equation}
We observe that also in this scenario, a large parameter space opens up for electron--muon non-universality that is consistent with the latest data on $R_K$.

\section{Conclusions}
\noindent In studies of rare $B$-meson decays, new sources of CP violation are often not considered. We have investigated the effects of such new sources and pointed out two interesting new features. First, we have shown how measurements of direct and mixing-induced CP violation in $B \to K\ell^+\ell^-$ decays can be used to extract the complex Wilson coefficients $C_{9\ell}$ and $C_{10\ell}$. Second, we have demonstrated how new sources of CP violation may open up a new parameter space for lepton flavour universality violation, allowing for significant violations of universality even with a value of $R_K$ in agreement with the SM prediction. Studies of CP violation constitute the next step in exploring the flavour sector of the Standard Model. These studies may lead to surprises and provide exciting new opportunities to reveal New Physics effects in the coming high-precision era.

\acknowledgments{A.R. would like to thank the organizers for the invitation to the enjoyable conference. This research has been supported by the Netherlands Organisation for Scientific Research (NWO). }

\clearpage
\bibliographystyle{JHEP} 
\bibliography{refs.bib}

\end{document}